\documentstyle[preprint,aps]{revtex}

\renewcommand{\slash}[2]{#1\kern-#2pt\mbox{\it/}}
\def\pp{{\bf p}}
\makeatother
\renewcommand{\thefootnote}{\alph{footnote}}

\begin{document}

\begin{center}
  \begin{Large}
    Polarization in $\bar{B}_c \rightarrow J/\psi~\mu~\bar\nu_{\mu}$
  \end{Large}
\end{center}
 
\vspace{2cm}

\renewcommand{\thefootnote}{\fnsymbol{footnote}}
\begin{center}
Myoung-Taek Choi\footnote{ e-mail:~mtchoi@hepth.hanyang.ac.kr}\\
Department of Physics, Hanyang University, Seoul 133-791,\\
\vspace{5mm}
Ji-Ho Jang   and Jae-Kwan Kim\\
Department of Physics, Korea Advanced Institute of Science and
Technology \\ 
373-1 Kusung-dong, Yusung-ku, Taejon 305-701, Korea.
\end{center}


\vspace{2cm}

\hspace{3mm}We study the polarization of the $J/\psi$ meson 
of the $\bar{B_c}\rightarrow J/\psi~\mu ~\bar \nu_{\mu}$ decay process,
followed by $J/\psi \rightarrow \mu^+ \mu^-$, 
with the help of the heavy-quark spin symmetry formalism (HQSS) of Jenkins
{\it et al.}~\cite{Jenk93} We adapt the ISGW wave function.
Due to the clean signature of the decay mode,
measurements of the polarization of the $J/\psi$ meson
can play a special role 
in extracting $|V_{cb}|$, the quark mixing-matrix element. 
We compare the results
with the predictions of other quark models. 

\vspace{5mm}
\noindent (PACS number: 13.25.Hw)

\newpage
\section{Introduction}

Semileptonic decays of pseudoscalar heavy mesons to vector mesons ($P
\rightarrow V$) provide richer physics than
those of pseudoscalar heavy mesons to pseudoscalar mesons ($P
\rightarrow P$).
First, the branching ratio of $P \rightarrow V$ is larger than that of
$P\rightarrow P$ due to the spin structure. 
Second, 
$B\rightarrow D^* ~l ~\bar\nu$ decay offers a
good chance to extract an accurate value of $|V_{cb}|$ because it is
less affected by $1/m_Q$ 
corrections~\cite{Neubert91}.
Finally, they allow a measurement of the polarization of the vector meson. 
The study of polarization of vector meson is important for the
following reasons: the degree of polarization of the
daughter particle, in general, strongly influences  the momentum
spectrum of the final particles as well as the decay rates; hence
the ratio of the longitudinal to the transverse decay width
is quite sensitive to form factors.
Also the extraction of individual form factors 
proves fruitful in determining $|V_{cb}|$~\cite{hagi89}.
Finally one can test quark models to get
a better understanding of the underlying structure and 
the dynamics of hadrons and currents.  

Several experiments have been performed to  measure the
$|V_{cb}|$~\cite{Vcb_meas} for the decay of $B$ mesons. 
Also, measurements of the form factors and the polarizations in 
$D \to K^*l\bar\nu$ and $B \to D^*l\bar\nu$ have been done~\cite{BD*pol}.
However, they suffer from large experimental uncertainties because of 
reconstruction due to cascade decays:
$D^*$'s are reconstructed from the decay chain $D~\pi$, followed by 
$D \rightarrow K~\pi$ or $K~\pi~\pi$.

Recently, much interest has been paid to the study of $B_c$ meson, 
which is a source of 
interesting physics; it provides a unique probe of both strong and
weak interactions. Unlike other heavy quarkonium systems which decay
strongly and electromagnetically, the $B_c$ meson can decay only
weakly because of the fact that it carries flavor explicitly. 
The decays of the $B_c$ meson can potentially provide a rich source 
for measuring the Cabibbo-Kobayashi-Maskawa (CKM) matrix elements. 
One also expects  rich spectroscopy for the $(\bar bc)$ bound 
states probing the inter-quark potential at distances intermediate to
the charm and the beauty quarkonium systems.
Candidates for $B_c \to J/\psi l \nu$ were recently reported for the
ALEPH detector at LEP~\cite{hwi}.
A quite large number of $B_c$ mesons are expected to be
produced by the future Large Hadron Collider (LHC) experiment.

In this paper, we study the production of 
the transverse and the longitudinal vector meson $J/\psi$ 
in semileptonic decays 
$\bar{B_c} \rightarrow J/\psi ~\mu ~\bar\nu_{\mu}$
by using the heavy-quark spin symmetry formalism (HQSS).
\setcounter{footnote}{0}
\renewcommand{\thefootnote}{\arabic{footnote}}
The exclusive decays of $B_c$ mesons, which include
the $J/\psi$ meson as a final state,
are essential to study  $B_c$ mesons
because the branching ratio is large and 
the $J/\psi$ meson decaying to dilepton pairs is easy to identify.
Among the exclusive decays,
the  $\bar{B_c}\rightarrow J/\psi ~\mu~\bar\nu_{\mu}$ process, 
followed by the decay of 
$J/\psi$ into a $\mu^+\mu^-$ pair,
will show clean experimental signatures\footnote{The electron
can also be used   instead of the muon. 
However, the signature of the muon is expected to be clearer than that of
an electron in huge backgrounds of hadronic collisions.}:
three energetic leptons coming from the  secondary vertex, two of them
reconstructing a $J/\psi$, and some missing transverse momentum 
due to neutrino.
Since the measurements of the polarizations of $J/\psi$ mesons
in $\bar{B_c} \rightarrow J/\psi ~\mu ~\bar\nu_{\mu}$ decays
can be used to constrain the form factor behavior with respect to
momentum transfer,
these measurement will play a special role in extracting the 
$|V_{cb}|$ matrix element. 

In Sec.~II,  we review the HQSS formalism of Jenkins {\it et
  al.}~\cite{Jenk93} and calculate the decay width of $\bar{B_c}
  \rightarrow J/\psi ~\mu ~\bar\nu_{\mu}$. 
The  polarization of the $J/\psi$ meson in $\bar{B_c}$ meson decay is
discussed in Sec. III.
We obtain the value of the longitudinal and the transverse decay
widths and find an expression for the polarization parameter which is
directly related to the measurement. 
In Sec.~IV, discussion and conclusions are given.
We compare the results to those of other quark models, such as
ISGW2 (the improved Isgur-Scora-Grinstein-Wise (ISGW)
model)~\cite{isgw2}, the original ISGW~\cite{isgw} 
and the  Bauer-Stech-Wirbel (BSW) model~\cite{bsw}. 
In the appendix,
we briefly summarize the formalisms of the ISGW and the BSW models we
used in the calculation.

\section{Heavy Quark Spin Symmetry of $B_c$ meson and semileptonic
  decay width}

The heavy-quark symmetry~\cite{IW} of Quantum Chromodynamics (QCD) in the
infinite quark mass limit has been successfully applied to the hadrons 
containing a single heavy quark.
In a  heavy-light quark system, such as the $B$ meson, the heavy quark act as
a static color source in the infinite-mass limit: light degrees of
freedom do not feel the change of a $b$-quark to a $c$-quark.
As a consequence, two symmetries occur; heavy-quark flavor symmetry
and heavy-quark spin symmetry \cite{IW}. Due to the two symmetries,
the form factors of heavy-meson decay are expressed in terms of a single
Isgur-Wise form factor, and the prediction of the matrix elements is
quite simplified.
In the $B_c$ meson, however,
both quarks should move around each other to make a stable meson.
Since the kinetic-energy term should be kept in the Lagrangian 
even at leading order, 
the flavor symmetry is broken explicitly.

Since the spin-spin interaction between the quarks is
proportional to $1/m_b m_c$ and is expected to be small, 
the spin symmetry still remains.
Jenkins {\it et al.}~\cite{Jenk93} investigated the consequences
of the heavy quark spin symmetry of the $B_c$ meson
and showed that the semileptonic 
decay $\bar{B_c} \rightarrow J/\psi~\mu~\bar\nu_{\mu}$ can be described
using only one form factor near zero recoil. 

Heavy-quark spin symmetry implies that the pseudoscalar $B_c$ meson is
degenerate with the vector $B_c^*$ meson.
The consequence of the spin symmetry of the heavy-heavy meson system
is compactly derived using the well-known covariant representation
formalism~\cite{trace}. 
The pseudoscalar $B_c$ meson of velocity $v$ is represented by a
$4\times 4$ matrix
\begin{equation}
  \label{bc-dbl}
  {\cal H}^{(\bar bc)} = \frac{(1+\slash{v}{5.8})}{2} [
  B_c^{*\mu}\gamma_{\mu} - 
  B_c \gamma_5 ] \frac{(1-\slash{v}{5.8})}{2},
\end{equation}
where $B_c$ and $B_c^*$ annihilate the pseudoscalar and the vector
meson $\bar bc$ bound states of velocity $v$, respectively.
Analogously, the $(\eta_c, J/\psi)$ spin multiplet of velocity $v'$
is given by 
\begin{equation}
  \label{cc-dbl}
  {\cal H}^{(\bar cc)} = \frac{(1+\slash{v'}{7.4})}{2}
  [J/\psi^{*\mu}\gamma_{\mu} - 
  \eta_c \gamma_5 ] \frac{(1-\slash{v'}{7.4})}{2}.
\end{equation}
The spin multiplet for $B_a$ and $B_a^*$ is given by
\begin{equation}
  \label{ba-dlb}
  {\cal H}_a^{(\bar b)} = [ B_a^{*\mu}\gamma_{\mu} -
  B_a \gamma_5 ] \frac{(1-\slash{v}{5.8})}{2},
\end{equation}
where the subscript $a=1,2,3$ (or u,d,s) is an $SU(3)_V$ flavor index.

The amplitudes for semileptonic $B_c$ decay to lower mass states
are determined by the matrix elements of the corresponding weak
hadronic current  between the meson states.
For example, the most general form for the matrix element of $B_c$
decay to $B_a$ and $B_a^*$ is
\begin{equation}
  \label{me-Ba}
  \langle B_a^{(*)},v,q\vert \bar q_a\Gamma c \vert B_c,v\rangle = 
  -\sqrt{m_{B_c}m_{B_a}}~{\bf tr}~({\cal H}_a^{(\bar b)}\Omega(v,q)
  \Gamma {\cal H}^{(c\bar b)}), 
\end{equation}
where 
\begin{equation}
  \Omega(v,q) = \Omega_1 + \Omega_2\slash{q}{5.8}
\end{equation}
is the most general Dirac matrix that can be written in terms of the
vectors $q$ and $v$. 
Explicit evaluation of Eq. (\ref{me-Ba}) gives
\begin{eqnarray}
  \label{me-spli}
  \langle B_a,v,q \vert V_{\mu} \vert B_c,v\rangle &=&
  2\sqrt{m_{B_c}m_{B_a}} [\Omega_1 v_\mu+\Omega_2 q_\mu ], \\ \nonumber
  \langle B_a^*,v,q \vert V_{\mu} \vert B_c,v\rangle &=&
  -2i\sqrt{m_{B_c}m_{B_a}}
  \Omega_2\epsilon_{\mu\nu\alpha\beta}\epsilon^{*\nu} q^\alpha v\beta,  \\
  \langle B_a,v,q \vert A_{\mu} \vert B_c,v\rangle &=&
  2\sqrt{m_{B_c}m_{B_a}} [\Omega_1 v_\mu+\Omega_2 \epsilon^*\cdot
  qv_\mu ], 
\end{eqnarray}
where $V_\mu$ and $A_\mu$ refer to the vector and the axial vector
currents, respectively, and $\epsilon_\mu$ is the polarization vector
of $B_a^*$.  
Here, six form factors are expressed in
terms of 2 independent form factors. 

One very different point with respect to single heavy-quark
systems is that the form factors are not normalized at the zero-recoil
point. The value depends on how exactly one calculates the bound-state
wave function.
Even in order to obtain an estimate of the corresponding decay width,
we need to extrapolate to the larger recoil region, which requires a
reasonable model.
The form factor $\Omega_2$ is irrelevant for $\mu$- or $e$-lepton
semileptonic decay  because the contribution of $\Omega_2$ to the decay
amplitude will be proportional to the lepton mass. In addition,
$\Omega_2$ does not contribute to decay amplitudes at zero recoil,
$q^2=0$.

In the case of $\bar{B_c} \rightarrow J/\psi$ decay, there is an
additional spin symmetry of the produced antiquark ($\bar c$), 
which forbids a form factor proportion to $q$.
The spin symmetry requires only one single form factor at the
zero-recoil point  for $\bar{B_c} \rightarrow
J/\psi~\mu~\bar\nu_{\mu}$~\cite{galdon}. 
The matrix element for the decay of $\bar{B_c}$ to $J/\psi$
is expressed by a single Isgur-Wise-like function $\Omega(v\cdot v')$,
which will be denoted as $\Delta$ hereafter:
\begin{equation}
  \label{me-gen}
  \langle J/\psi,v'\vert \bar c\Gamma b \vert B_c,v\rangle = 
  -\sqrt{m_{B_c}m_{J/\psi}}~\Delta(v\cdot v')~{\bf tr}~({\cal
  H}^{(\bar bc)} \Gamma {\cal H}^{(\bar  cc)}), 
\end{equation}
where 
$\epsilon_{\mu}$ is the polarization vector of the $J/\psi$.
From the above formula, one finds~\cite{Jenk93}
\begin{eqnarray}
  \label{me}
  \langle J/\psi, v' \vert V_{\mu} \vert B_c,v\rangle &=&  0, \\
  \label{me2}
  \langle J/\psi, v' \vert A_{\mu} \vert B_c,v\rangle &=& 
  2 \sqrt{m_{B_c}m_{\eta_c}} \Delta(v\cdot v') \epsilon_{\mu}^*,
\end{eqnarray}
where $V_{\mu}$ and $A_{\mu}$ refer to the vector and the axial vector
currents respectively.

In the limit of vanishing lepton mass, we finds the differential decay
width:
\begin{eqnarray}
  \label{dif_delta}
  \frac{d\Gamma}{d\omega} = \frac{G^2_F}{48 \pi^3}\vert V_{cb} \vert^2
  m^{2}_{B_c} m^3_{J/\psi} (\omega^2-1)^{1/2} F(r,\omega)
\end{eqnarray}
with
\begin{eqnarray}
  F(r,\omega)~ =& 8(1-2\omega r +r^2)\Delta^2, 
   &~~~B_c \rightarrow J/\psi_T ~\mu~\bar\nu_{\mu}, \\
  F(r,\omega)~ =& 4(\omega-r)^2 \Delta^2,  &~~~B_c \rightarrow
   J/\psi_L ~\mu~\bar\nu_{\mu}.  
\end{eqnarray}
In the above equation,
\begin{equation}
  w = v \cdot v' = \frac{m^2_{B_c}+m^2_{J\psi}-q^2}{m_{B_c}+m_{J/\psi}} 
\end{equation}
and $r = m_{J/\psi}/m_{B_c}$, where $q^2$ is the momentum transfer to
the lepton pair.

The heavy-quark symmetry itself does not predict the exact structure
of the form factor $\Delta(\omega)$,
which includes all the non-perturbative dynamics
of the system.
Ref. 1 showed that by using the operator product expansion,
$\Delta(q)$ could be expressed by wave function overlaps
of the initial and the final mesons
\begin{equation}
  \label{overlap}
  \Delta(q)  = \int d^3x~\textrm{e}^{-i\vec{q}\cdot\vec{x}}
  \Psi^*_{J/\psi}(x) \Psi_{B_c}(x).
\end{equation}

We adopt the non-relativistic wave function of the ISGW
model~\cite{isgw} to estimate the polarization of the $J/\psi$.
$\Delta(\omega)$ is expressed as,
\begin{equation}
  \label{delta}
  \Delta(\omega) =
  \left (\frac{2\beta_{B_c}\beta_{J/\psi}}{\beta_{B_c}^2+
      \beta_{J/\psi}^2}\right )^{3/2} 
  \exp \left (-\frac{m^2_{sp}}{\kappa^2(\beta_{B_c}^2+\beta_{J/\psi}^2)}
  (\omega-1)\right),
\end{equation}
where $\beta_{B_c}$ and $\beta_{J/\psi}$ are  parameters of the model,
$m_{sp}$ is the mass of the spectator quark,
and $\kappa$ is introduced to account for the relativistic recoil
effect.  
Since $\beta_{B_c}$ is not the same as $\beta_{J/\psi}$,
the form factor $\Delta(\omega)$ is not normalized to unity at the
zero-recoil point, which reflects breaking of the heavy flavor symmetry. 
The resulting decay width is shown in Table 1
for $\kappa = 0.7$ (the value used in the Ref. 9).

\section{Polarization of $J/\psi$ meson}

In $B_c \to J/\psi l \bar\nu$ decay, followed by $J/\psi$ decaying into two
lepton pairs, the decay amplitude with  helicity $\lambda$ for
$J/\psi$  is~\cite{hagi89,ks90} 
\begin{equation}
  \label{eq:vll}
  {\cal M} \propto \frac{G_F}{\sqrt{2}} V_{cb} \sum_{\lambda}
  L_\lambda H_\lambda d^1_{\lambda_{J/\psi},\lambda_{l^-}-\lambda_{l^+}}, 
\end{equation}
where $\lambda_{l^-}-\lambda_{l^+} = \pm 1$.
Here, $L_\lambda$ describe the $W_\lambda^{*-} \to l\bar\nu$ decay,
$H_\lambda $ are the three $\bar B \to D_\lambda^* W_\lambda^{*-}$
decay amplitudes, and the Wigner $d$-function describes the vector
meson decay into two fermion pairs, $J/\psi \to l^+l^-$.

The angle distribution of $l^-$ in the rest frame of $J/\psi$ is 
given by
\begin{equation}
  \frac{d\Gamma}{d \cos\theta_{l^-}} \propto 1 + \alpha'
  \cos^2\theta_{l^-},  
\end{equation}
where we find the polarization parameter $\alpha'$ is expressed
\begin{equation}
  \label{polpar}
  \alpha' = \frac{\Gamma_T - 2\Gamma_L}{\Gamma_T + 2\Gamma_L}.
\end{equation}
In the case of a vector meson decaying into two pseudoscalar mesons,
like $D_\lambda^* \to D\pi$ decay, the Wigner function should
be replaced by $J=1$ spherical harmonics  $Y^1_\lambda$. Then, the
polarization parameter is of the familiar form
\begin{equation}
  \label{polparB}
  \alpha = 2\frac{\Gamma_T}{\Gamma_L} -1.
\end{equation}

We note that
the  parameter $\alpha'$ is expressed as the ratio of  the
longitudinal and the transverse decay widths, as shown in the 
Eq.~(\ref{polpar}). As we saw in the previous section, $\Delta(1)$ is
not normalized to 1 due to the breaking of the heavy-flavor symmetry,
and a model is required to predict the value. 
However, $\Delta(1)$ is canceled in Eq.~(\ref{polpar}); therefore, it is
irrelevant to $\alpha'$.
As a result, the right-hand side of Eq.~(\ref{polpar}) 
is only a function of the slope parameter.
One cannot predict the slope of the form factor at the
non-zero recoil point, which is essentially
non-perturbative, without referring to the quark-model, even in a
heavy-light meson system like $B$.
The experimental measurement of $\alpha'$  can be used to
constrain the value of the slope parameter in the semileptonic decay
of $B_c$ meson, but  we will  have to wait for this 
until sufficient data  are gathered in the future.

Here, we evaluate the value of $\alpha'$ using  Eq.~(\ref{delta}).
Integrating  Eq.~(\ref{dif_delta}) over the possible kinetic-energy  range,
the following result is obtained (the parameters we used are shown in
the Table 1):
\begin{eqnarray}
  \Gamma_L &=& 9.6 \times 10^{-15} ~ GeV, \\
  \Gamma_T &=& 7.0 \times 10^{-15} ~ GeV, \\
  \frac{\Gamma_L}{ \Gamma_T} &=& ~1.37, \\
  \alpha  &=& -~0.47.
\end{eqnarray}
Fig. 1 shows the plot of the transverse and the longitudinal decay
widths with respect to $\omega$ for $\kappa = 0.7$.
$\Gamma_L$ dominates at low $q^2$ (high $\omega$): at $q^2$ = 0,
$J/\psi$ has its maximum possible momentum, and the helicities are
aligned to give $S_z$ = 0.
$\Gamma_T$ dominates near $q^2=q^2_{\textrm{max}}$(low $\omega$):
at small $J/\psi$ velocity  the probabilities are uncorrelated with
the spin of $J/\psi$; hence, $H_+ = H_-= H_L$. This gives
$\Gamma_T/\Gamma_L = 2$. 

The authors of Ref. 9 obtained the value of the correction
factor $\kappa$ from the pion form factor by 
comparison with  experiment.
In our case, however, there are no available data to determine the
correct value of $\kappa$. Therefore,
we varied $\kappa$ from 0.6 to its maximum of 1 and 
Fig. 2 shows  $\alpha'$ as a function of the
relativistic correction factor $\kappa$. Within the assumption of
the validity of Eqs.~(\ref{me}) and (\ref{me2}) measuring $\alpha'$
determines $\kappa$.

\section{Discussion and Conclusions}

Polarizations of vector meson have been studied by several
authors~\cite{hagi89,IS-pol,bswpol,gil90,ks}.
We compare the above predictions with those of other quark models which have
been generally regarded as giving a successful description of the
semileptonic decays of heavy mesons like the $B$ meson.
The result is given in the  Table 1.
We find that $\Gamma_L/\Gamma_T$ is rather large in the heavy-quark
spin symmetry formalism when the ISGW model is adopted.

We  consider some uncertainties of our method. 
First, strictly speaking, the expressions of Eq.~(\ref{me}) are valid
near the zero-recoil point. Additional form factors might contribute at
the large recoil point. The exact expression of the matrix elements at
large recoil within the heavy-spin symmetry formalism
is beyond the scope of this paper.
However, since the recoil momentum of $J/\psi$ is small
($\omega-1 \simeq 0.26$) due to its heavy mass,  
we expect that the assumption of Eq.~(\ref{me}) being applicable to other
kinematic points is not too wrong. 
Second, in order to obtain an estimate of the decay width,
we need to extrapolate the form factor to  the large-recoil region,
which requires a 
reasonable model. Since we used the ISGW model wave function in the 
calculation of the form factor, model dependences can not be avoided.
The value of the  polarization is also sensitive to 
the parameter $\kappa$. Unfortunately, one cannot say which value of
$\kappa$ should be used for the form factor;
experiments should provide that information.
We checked that  $\Gamma_L/\Gamma_T$ is insensitive to the
parameter set, such as quark mass. 
We note that even in $B\to D^*l\nu_l$, there are large
discrepancies in the $\Gamma_L$ and $\Gamma_T$
between the heavy-quark effective theory (HQET) and other models, 
although they all are consistent within the experimental error.

Finally, we remark that
the measurement of the polarization of the $J/\psi$ meson
in the $\bar{B_c} \rightarrow J/\psi ~\mu ~\bar\nu_{\mu}$ decay
may provide an alternate way to extract $|V_{cb}|$.
Knowing both the form factor behavior on the right-hand side of 
Eq.~(\ref{dif_delta}) and the measured value of the total decay width 
on the left-hand side allows us to determine the value of $|V_{cb}|$.
As we saw in Eq.~(\ref{overlap}), $\Delta(1)$ is a product of two meson
wave functions at rest. 
Although the evaluation of the wave function of the meson at rest
depends on the model, the uncertainty is expected to be small
compared to that of a non-zero recoil meson. 
Even the value of $\Delta(1)$ can be obtained from lattice
QCD with small uncertainty.
In the case of $B \to D^*l\nu$ decay, $\Delta(1)$ is normalized to one as a
consequence of HQET. Hence, the measurement of the slope parameter through
the polarization of $D^*$ (Eq.~(\ref{polparB})) and the measurement of
the decay width determines the  $|V_{cb}|$ model independently.
Further discussions will be published elsewhere~\cite{vcbpol}.

\begin{center}
  \bf ACKNOWLEDGMENTS
\end{center}

We would like to thank K. Y. Lee for helpful
discussions and for reading the manuscript. This work was supported 
by the Korea Research Foundation and by the Korea Science and 
Engineering Foundation.

\appendix

\section{ISGW model}

The meson state vector is expressed by the following nonrelativistic
expression:
\begin{eqnarray}
  |X(\pp_X; s_x)> &=& \sqrt{2m_x}\int d^3p \sum C_{m_Lm_S}^{s_xLS}
  \phi_X(\pp)_{Lm_L} \chi_{s\bar s}^{Sm_S}  \\ \nonumber
  && |q(\frac{m_q}{m_x} \pp_X + \pp,s) \bar q (\frac{m_{\bar q}}{m_x} 
  \pp_X - \pp, \bar s) >,
\end{eqnarray}
where $\chi_{s\bar s}^{Sm_S}$ is the spin wave function of the
quark-antiquark pair in the state with  total spin $S$ and  spin
projection $m_S$; $C_{m_Lm_S}^{s_xLS}$ is the coupling between the
orbital momentum $L$ and the total spin $S$ of a system with the
total momentum $s_x$; $\phi_X(\pp)_{Lm_L}$ is the corresponding
nonrelativistic wave function; $\pp_X$ is the meson momentum; and $\pp$ is
the relative momentum of quarks. In the model, the meson mass is equal
to the sum of quark masses. As the probe wave functions, the
nonrelativistic oscillator wave functions are used.

In the ISGW model,  the hadronic matrix elements are defined as 
\begin{eqnarray}
  \label{ff_isgw}
  <m(k)\mid V_\mu(0)\mid M(P)> &=& f_+(q^2)(P+k)_\mu +
  f_-(q^2)(P-k)_\mu, \\
  <m(k,\epsilon^*)\mid V_\mu(0)\mid M(P)> &=&
  ig(q^2)\epsilon_{\mu\nu\rho\sigma}\epsilon^*(P+k)^\rho(P-k)^\sigma,\\
  <m(k,\epsilon^*)\mid A_\mu(0)\mid M(P)> &=&
  f(q^2)\epsilon^*+a_+(q^2)(\epsilon^*\cdot P)(P+k)_\mu \\ \nonumber
  && + a_-(q^2)(\epsilon^*\cdot P)(P-k)_\mu 
\end{eqnarray}
Expressions of the form factors 
were also obtained (see Refs. 9 and 15).
We define
\begin{eqnarray}
  \bar{H}_{\pm} &=& [ f(q^2) \mp 2 M K g(q^2)], \\
  \bar{H}_0     &=& \frac{M}{2m\sqrt{y}} [(1-\frac{m^2}{M^2}-y) f(q^2)
  + 4 K^2 a_+(q^2) ].
\end{eqnarray}
For the transition into a pseudoscalar meson final state, one obtains
\begin{equation}
  \bar{H}_{\pm} = 0, ~\bar{H}_0 = -2\frac{K}{\sqrt{y}}f_+(q^2).
\end{equation}
In this case,
\begin{equation}
   \frac{d\Gamma}{d\omega} = ~\frac{G^2_F\mid V_{Qq}\mid^2 K^3 M^2}{24\pi^3}
   \mid f_+(q^2)\mid^2.
\end{equation}
For the vector-meson final states, in the $l\nu$ frame
\begin{equation}
  {\bf H} = 2i\sqrt{\omega}
  Mg(q^2) {\bf \epsilon}^*\times\ {\bf k}
  -f(q^2){\bf \epsilon}^* - 2(\epsilon^*\cdot P)a_+(q^2)
  {\bf k} ,
\end{equation}
where $a_-$ does not contribute. Then
\begin{equation}
  \frac{d\Gamma}{d\omega} = ~\frac{G^2_F\mid V_{Qq}\mid^2 K
    M^2\omega}{96\pi^3} 
  [ \mid\bar{H}_+\mid^2 +
  \mid\bar{H}_-\mid^2 + \mid\bar{H}_0\mid^2 ].
\end{equation}

\section{BSW model}

The meson is considered as a relativistic bound state of a quark
$q_1$ and an antiquark $\bar q_2$ in a  system of infinitely large
momentum:
\begin{eqnarray}
  |P,m,j,j_z> &=& \sqrt{2}(2\pi)^{3/2}\sum_{s_1,s_2} \int ~d^3p_1~d^3p_2
  \delta^3 (\pp-\pp_1-\pp_2) \\ \nonumber
    && L_{m}^{j,j_z}(\pp_{1t},x,s_1,s_2) a_1^{{s_1}\dagger}(\pp_1)
    b_2^{{s_2}\dagger} (\pp_2) |0>,
\end{eqnarray}
where $P_\mu=(P,0,0,P)$; as $P\to\infty$, $x=p_{1z}/p$ corresponds
to the momentum fraction carried out by the nonspectator quark; 
and $p_{1t}$ is the transverse momentum.
For the orbital part of the wave functions, the solution of the
relativistic oscillator is used
\begin{eqnarray}
  L_m(\pp_t,x) &=& N_m \sqrt{x(1-x)} \exp(-\frac{\pp_t^2}{2\omega^2}) \\
  && \exp[ -\frac{m^2}{2\omega^2}
  (x-\frac{1}{2}-\frac{m_{q_1}^2 -m_{q_2}^2}{2m^2} )^2].\nonumber
\end{eqnarray}

The set of (dimensionless) form factors of the BSW  model is related to
that of the ISGW model: 
\begin{eqnarray} 
  F_1 (q^2) &=&  f_+ (q^2),   \\ \nonumber
  V   (q^2) &=& (M+m) g (q^2), \\ \nonumber
  A_1 (q^2) &=& (M+m)^{-1} f (q^2), \\ \nonumber
  A_2 (q^2) &=& -(M+m) a_+ (q^2).
\end{eqnarray}
One finds
\begin{eqnarray}
  H_{\pm}(q^2) &=& (M+m)A_1(q^2) \mp 2 \frac{MK}{M+m}V(q^2), \\
  H_0(q^2)     &=& \frac{1}{2m\sqrt{q^2}}\cdot 
        [(M^2-m^2-q^2)(M+m)A_1(q^2)-4\frac{M^2K^2}{M+m}A_2(q^2)]
\end{eqnarray}
with
\begin{equation}
  \nonumber
  K=\frac{1}{2M}[(M^2-m^2-q^2)^2-4m^2q^2]^{1/2}.
\end{equation}


\newpage

\begin{center}
  \large{FIGURES}
\end{center}

1. Plot of the transverse and the longitudinal decay widths
as functions of  $\omega = v\cdot v' $ (in units of $10^{-15}$ GeV).

\vspace{2cm}

2. Plot of $\alpha'$ as a function of  $\kappa$

\newpage

\begin{center}
  \large{TABLES}
\end{center}

Table 1. The widths ($10^{-15}$ GeV) of 
$ \bar{B_c}\rightarrow J/\psi~\mu~\bar\nu_{\mu} $, and
$\Gamma_L/\Gamma_T$     of the $J/\psi$ meson:
first row, heavy-quark spin symmetry~\cite{Jenk93};
second row, the ISGW model~\cite{isgw};
third row, values from Ref. 8;
fourth row, the BSW model~\cite{bswpol}.
We set $V_{cb} = 0.04, ~m_{B_c}=6.3~GeV, ~m_b=4.75~GeV, ~m_c=1.5~GeV,
~\kappa = 0.7$. Also,
$\beta_{B_c} = 0.82, ~\beta_{J/\psi}=0.61, ~\omega_{B_c}=0.8 ~GeV$,
$\omega_{J/\psi}=0.6 ~GeV$, as  used in Ref. 19.

\begin{center}
  \begin{tabular}{|l|cc|} \hline
    Model & $\Gamma$ & $\Gamma_L / \Gamma_T$  \\ \hline
    Here   & 16.6   & 1.37  \\
    ISGW  & 12.4   & 0.73  \\
    ISGW2 & 15.7   & 0.87  \\ 
    BSW   & 19.0   & 1.04  \\ \hline
  \end{tabular}
\end{center}

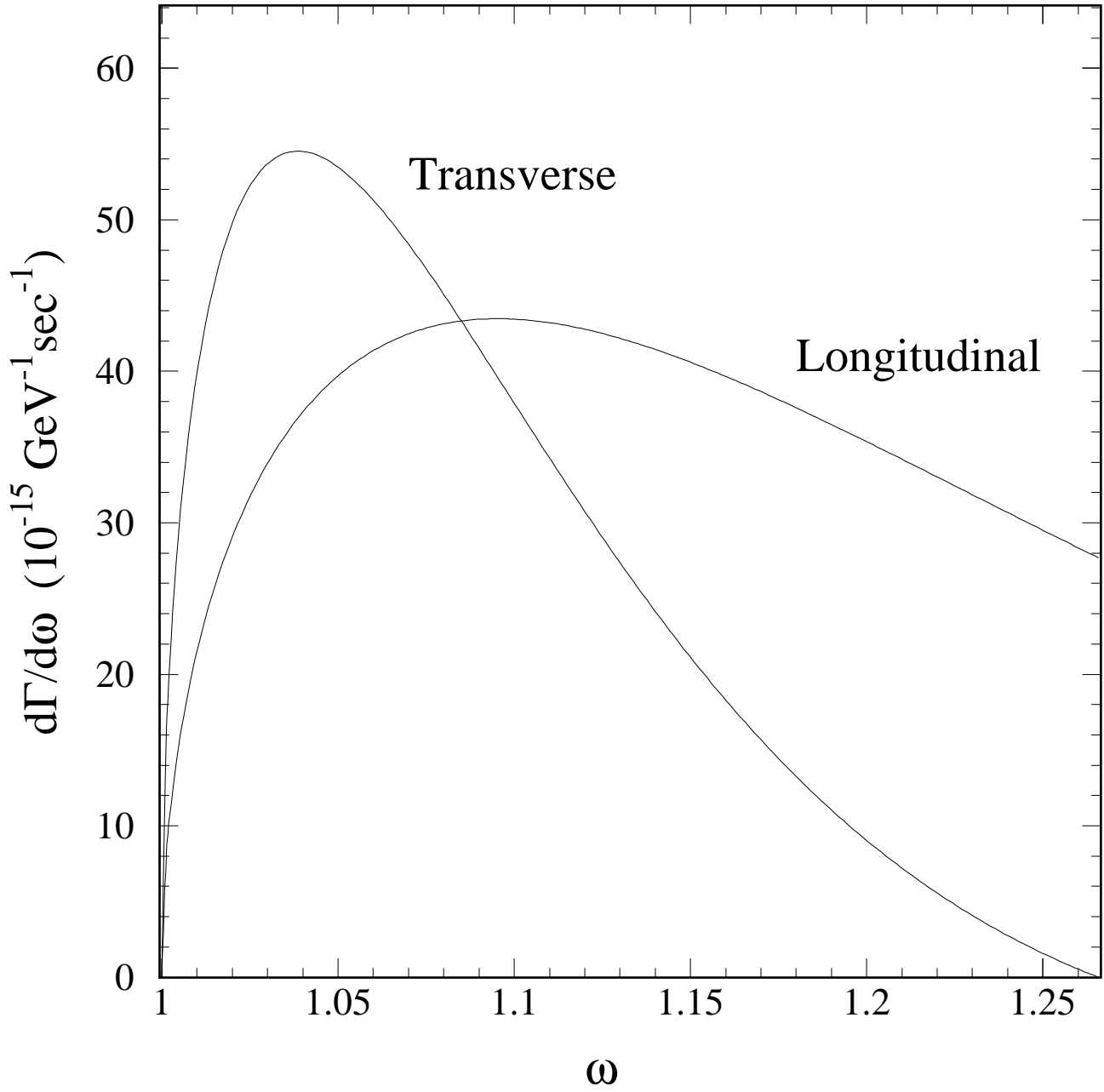
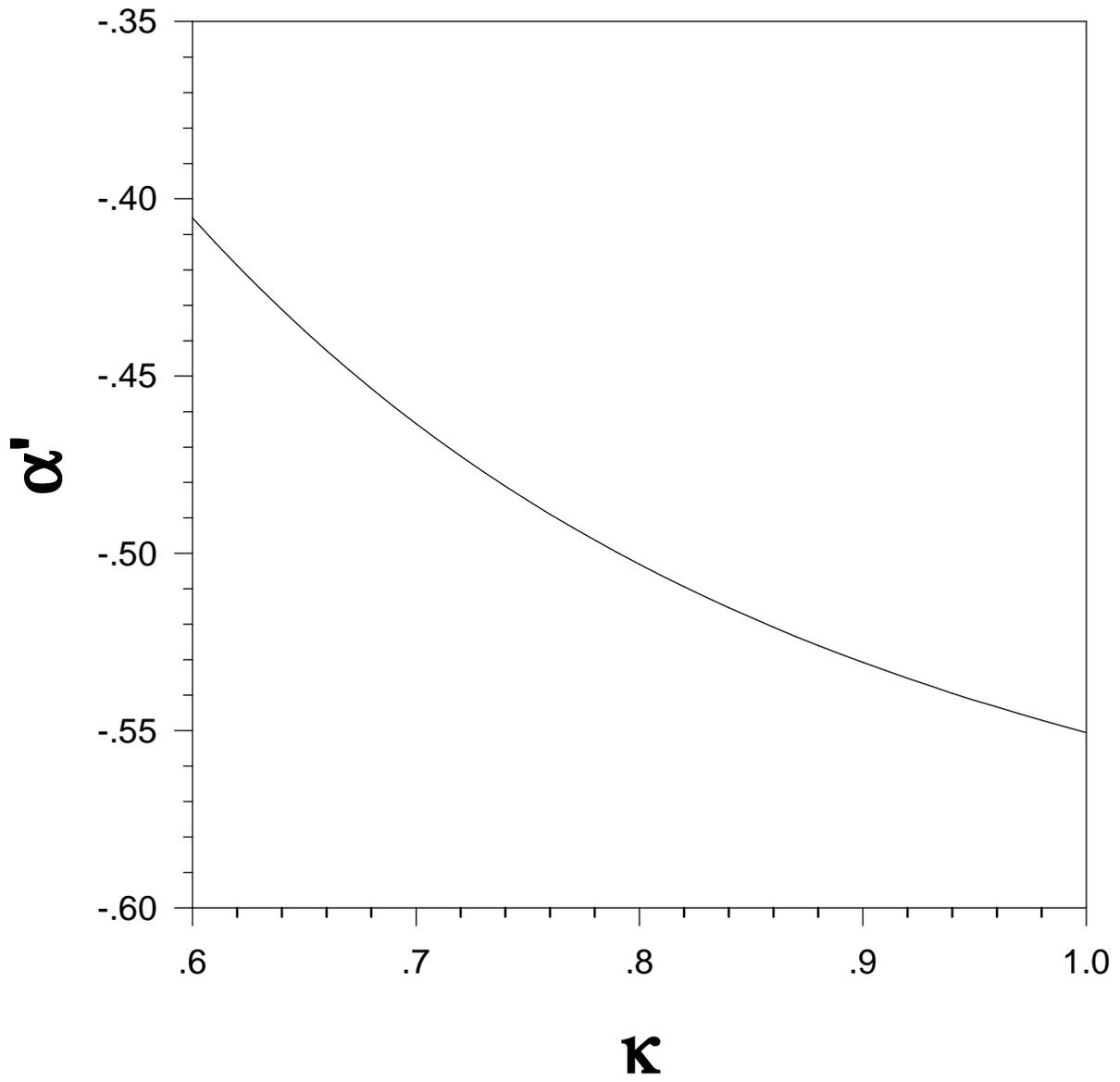
\end{document}